\documentclass[onecolumn,authoryear]{els-mrw} 
\usepackage{amsmath,amssymb,amsfonts,amsthm,bm,makeidx,graphicx}
\usepackage{txfonts}
\usepackage{helvet}
\usepackage{relsize}
\usepackage{mathrsfs}
\usepackage[authoryear,round]{natbib}  % or square, numbers, etc.
\newcommand{\kFermi}[1]{k_{\raisebox{-0.5pt}{\scriptsize{\rm \hspace{0.4pt}F}}}^{\raisebox{0.3pt}{\scriptsize #1}}}
\newcommand{\eFermi}[1]{\epsilon_{\raisebox{-0.5pt}{\scriptsize{\rm \hspace{1pt}F}}}^{\raisebox{0.3pt}{\scriptsize #1}}}
\newcommand{\xFermi}[1]{{\mathlarger x}_{\raisebox{-0.5pt}{\scriptsize{\rm \hspace{0.4pt}F}}}^{\raisebox{0.3pt}{\scriptsize #1}}}
\newcommand{\yFermi}[1]{y_{\raisebox{-0.5pt}{\scriptsize{\rm \hspace{0.4pt}F}}}^{\raisebox{0.3pt}{\scriptsize #1}}}
\newcommand{\pFermi}[1]{p_{\raisebox{-0.5pt}{\scriptsize{\rm \hspace{0.4pt}F}}}^{\raisebox{0.3pt}{\scriptsize #1}}}

\newcommand{\xFstar}[1]{{\mathlarger x}_{\raisebox{-0.5pt}{\scriptsize{\rm \hspace{0.4pt}F}}}^{\raisebox{0.3pt}{\scriptsize$\star$}}}
\newcommand{\yFstar}[1]{{\mathlarger y}_{\raisebox{-0.5pt}{\scriptsize{\rm \hspace{0.4pt}F}}}^{\raisebox{0.3pt}{\scriptsize$\star$}}}
\newcommand{\rhozero}{\rho_{\raisebox{-1.0pt}{\tiny\!0}}}
\newcommand{\epszero}{\varepsilon_{\raisebox{-0.5pt}{\tiny 0}}}

\begin{document}

\chapter{Relativistic mean-field models of neutron-rich matter}\label{chap1}

\author[1]{J. Piekarewicz}%

\address[1]{\orgname{Florida State University}, 
                    \orgdiv{Department of Physics}, 
                    \orgaddress{Tallahassee, Florida 32306, USA}}
                    
%\articletag{Chapter Article tagline: update of previous edition,, reprint..}

\maketitle

%\begin{glossary}[Glossary]
%\term{Europe} the model is a coherent view of capital markets data that allows users to interact with the content in a consistent manner.
%\term{Primates} regardless of the source. Essentially, of sources. Properly deployed.
%\end{glossary}

%\begin{glossary}[Nomenclature]
%\begin{tabular}{@{}lp{34pc}@{}}
%AF &Assessment Factor\\
%ECHA &European Chemical Agency\\
%EPM &Equilibrium Partitioning Method Equilibrium Partitioning Method Equilibrium Partitioning Method Equilibrium\hfill\break %Partitioning Method\\
%ERA &Ecological Risk Assessment\\
%HC &Hazardous Concentration\\
%\end{tabular}
%\end{glossary}

\begin{abstract}[Abstract]
 The aim of this chapter, focused on relativistic mean-field models and part of the Encyclopedia of Nuclear Physics, is 
 to provide an introductory, self-contained discussion accessible to a broad audience, including advanced undergraduate 
 students. The chapter surveys the fundamental ideas, assumptions, and theoretical framework underlying relativistic 
 mean-field models, and illustrates their wide range of applications across nuclear science. Particular emphasis is placed 
 on the central role that these models play in the construction of equations of state for strongly interacting matter, as well 
 as on the intimate connections between nuclear experiments, astrophysical observations, and theoretical modeling. In this 
 context, relativistic mean-field theory is shown to provide a unified description of bulk nuclear properties and dense 
 neutron-rich matter, enabling the interpretation of the remarkable structural and observational properties of neutron stars 
 in the emerging era of multi-messenger astronomy.
\end{abstract}

%%%%%%%%%%% Key Points %%%%%%%%%
\noindent\textbf{Key points}
\begin{itemize}
\item \textbf{Theoretical Foundation:} Relativistic mean-field (RMF) models utilize a framework where nucleons 
                     interact via the exchange of scalar and vector mesons. 
\item \textbf{Equation of State:} A thermodynamic relation that quantifies how the pressure evolves as a 
                    function of the energy density. Relativistic mean-field models are critical for constructing equations 
                    of state to describe the structure of neutron stars.
\item \textbf{Symmetry Energy:} The nuclear symmetry energy quantifies the energy cost of converting symmetric 
                    nuclear matter into pure neutron matter---an essential component of describing neutron stars.
\item \textbf{Nuclear Saturation:} The models provide a natural explanation for the saturation of symmetric nuclear 
                    matter, which occurs due to the competition between intermediate-range scalar attraction and short-range 
                    vector repulsion.
\item \textbf{Astrophysical Constraints:} Modern RMF functionals are calibrated against both laboratory data and 
                    astrophysical observations, such as the monitoring of heavy pulsars and gravitational wave data.                    
\item \textbf{Chemical Equilibrium:} In neutron stars, the RMF framework incorporates charge neutrality and beta 
                    equilibrium, predicting a proton-to-neutron ratio that approaches 1/8 at high densities. 
\item \textbf{Neutron Stars:} Neutron stars serve as astrophysical laboratories where theoretical models are tested 
                    at extreme densities and neutron-proton asymmetries. When coupled to the Tolman--Oppenheimer--Volkoff 
                    equations, these models yield predictions for neutron-star masses, radii, and tidal deformabilities that 
                    can be directly compared with modern multimessenger observations.
\end{itemize}
%%%%%%%%%%%%%%%%%%%%%%%%%%

\section{Introduction}
\label{sec:Intro}

Among the most compelling questions animating nuclear astrophysics today are: What novel states of matter emerge 
under conditions of extreme density and temperature? How were the heavy elements, from iron to uranium, forged 
in the cosmos? These---and many other fundamental questions---are addressed in the most recent Long Range Plans 
for Nuclear Science\,\citep{LRP2015,LRP2023}. Developed by the entire U.S. nuclear physics community, the long range 
plans chart a vision for the future and highlight the unparalleled breadth and depth of the field, spanning phenomena 
from the quark-gluon substructure of the nucleon to the microphysics of gigantic stellar explosions.

The exotic environments that shape nuclear astrophysics, from the synthesis of the heavy elements, to the composition 
of neutron stars, and to the cataclysmic mergers of binary neutron stars involve the presence of neutron-rich matter. 
Remarkably, in one clean sweep, the historical detection of gravitational waves from the binary neutron star merger 
GW170817 by the LIGO-Virgo collaboration\,\citep{Abbott:PRL2017} has provided critical insights into the nature of dense 
matter\,\citep{Bauswein:2017vtn,Fattoyev:2017jql,Annala:2017llu,Abbott:2018exr,Most:2018hfd,Tews:2018chv,Malik:2018zcf,
Radice:2017lry, Radice:2018ozg,Tews:2019cap,Capano:2019eae,Tsang:2019mlz,Tsang:2020lmb,Drischler:2020hwi,
Landry:2020vaw,Xie:2020rwg,Essick:2021kjb,Chatziioannou:2021tdi} as well as on the synthesis of the heavy elements 
in the cosmos\,\citep{Drout:2017ijr,Cowperthwaite:2017dyu,Chornock:2017sdf,Nicholl:2017ahq}.

Combined with this historic discovery, long-term pulsar-timing observations have determined with high precision the mass 
of the millisecond pulsar PSR~J0740+6620\,\citep{Cromartie:2019kug,Fonseca:2021wxt}. With a measured mass of 
$M\!=\!2.08\pm0.07\,M_{\odot}$, this compact object is currently among the most massive---and precisely measured---neutron 
stars known to date. By itself, this measurement already places stringent constraints on the equation of state, as it implies that 
the pressure support against gravitational collapse at the highest densities reached in the stellar interior must be sufficiently 
large to sustain such a massive star.

In turn, the Neutron Star Interior Composition Explorer (NICER) relies on  pulse-profile modeling of X-ray emission from 
localized hot spots on the stellar surface to enable the simultaneous determination of neutron-star masses and radii. To 
date, among the sources that have been targeted are: PSR~J0030+0451\,\citep{Riley:2019yda,Miller:2019cac}, the 
aforementioned millisecond pulsar PSR~J0740+6620\,\citep{Riley:2021pdl,Miller:2021qha}, and the brightest rotation-powered 
pulsar PSR~J0437-4715\,\citep{Choudhury:2024xbk}. These pioneering observations---and future measurements with enhanced 
sensitivity---are of critical importance in the determination of the equation of state of neutron-star matter. Indeed, the simultaneous 
determination of masses and radii provides one of the most direct avenues for constraining the nuclear equation of 
state via the inverse stellar structure problem\,\citep{Lindblom:1992,Lindblom:2025kjz}.

With the goal of providing an introductory and self-contained overview of relativistic mean-field models of nuclear matter, 
this chapter is written for a broad readership that includes both undergraduate and graduate students. The presentation 
emphasizes relativistic mean-field theory as a foundational framework for understanding the nuclear equation of state and 
its wide-ranging applications in nuclear physics and astrophysics, particularly in the description of neutron-star matter. To 
this end, the formal development begins with a discussion of the basic concepts underlying the equation of state, introduced 
in the simplified setting of relativistic free Fermi gases. In this context, a one-component Fermi gas is examined in detail, 
with all important results presented in closed analytic form. Recognizing that nuclear systems are composed of both neutrons 
and protons, the formalism is then extended to two-component systems, leading naturally to the introduction of the nuclear 
symmetry energy. The discussion is further generalized to incorporate the conditions of charge neutrality and chemical equilibrium 
that are essential for describing neutron-star matter. With all these key concepts in place, the relativistic Walecka model is 
introduced to illustrate how fundamental nuclear properties---such as saturation of symmetric nuclear matter---emerge naturally 
from the relativistic structure of the theory. Finally, representative modern extensions of the Walecka model are discussed as 
examples of modern energy-density functionals calibrated, with quantified theoretical uncertainties, to a broad set of laboratory 
and astrophysical observables.

\section{Formalism}
\label{sec:Fromalism}

Before venturing into the nuclear domain---where strong interactions are essential to 
unravel the complex dynamics of nuclear matter---we first introduce a simple ``toy model" 
that illustrates the critical role of quantum mechanics in shaping the equation of state of a 
free Fermi gas. A free Fermi gas is an idealized collection of identical fermions that do not 
interact with one another. In the quantum regime relevant to such systems, the temperature 
is low and the density is high. Hence, the only form of ``communication" between fermions 
arises from the Pauli exclusion principle. That is, the occupation of a single-particle state 
with energy $\varepsilon({\bf k})$ is governed by the Fermi-Dirac distribution:
%%%%%%%%%%%%%
\begin{equation}
 n(\varepsilon) = \frac{1}{1+\large{e}^{\,\beta(\varepsilon-\mu)}},
 \label{FermiDirac} 
\end{equation}
%%%%%%%%%%%%%
where $\varepsilon({\bf k})$ is the energy of a (non-interacting) single-particle state, ${\bf k}$
is the quantized momentum associated to that state, $\beta\!=\!(k_{\rm B}T)^{-1}$ is the inverse
temperature in units of the Boltzmann constant, $T$ is the temperature, and $\mu$ is the
chemical potential. For a system containing $N$ fermions, the chemical potential serves as 
a  ``Lagrange multiplier" that is adjusted to ensure that the total number of particles equals 
$N$. 

%%%%%%%%%%%%%
\begin{figure}[ht]
\centering
\includegraphics[width=0.6\textwidth]{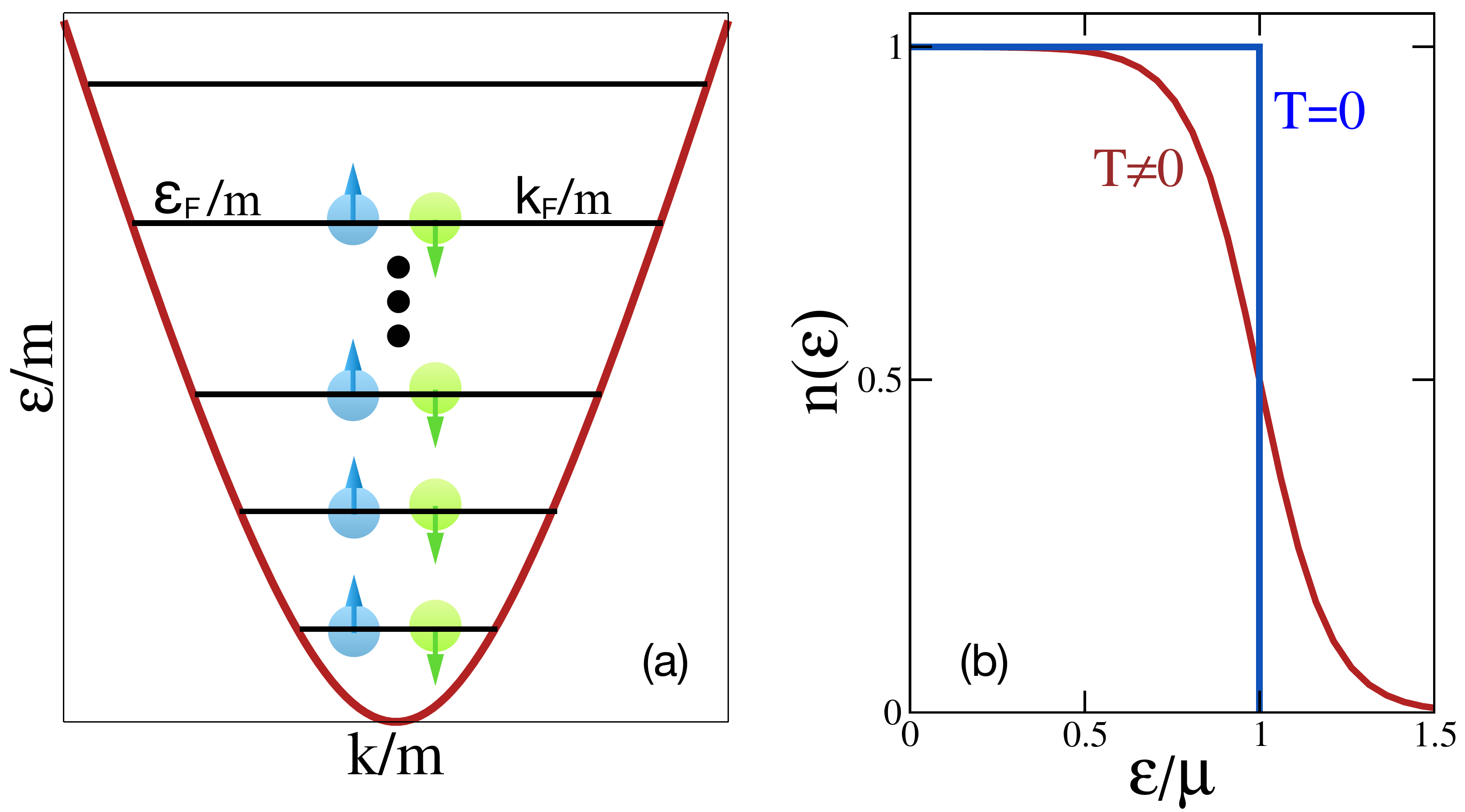}
\caption{Single-particle properties of a relativistic free Fermi gas. (a) Single-particle energy 
$\varepsilon/m$ as a function of the momentum $k/m$, illustrating the relativistic dispersion relation. 
The Fermi momentum $k_F$ and the corresponding Fermi energy $\varepsilon_F$ are indicated at
the last occupied single-particle state. All states above that are empty. (b) Occupation probability 
$n(\varepsilon)$ as a function of the scaled single-particle energy $\varepsilon/\mu$, where $\mu$ 
denotes the chemical potential. At zero temperature ($T\!=\!0$), the distribution reduces to a sharp step 
function, whereas at finite temperature ($T\!\neq\!0$) thermal effects smear the Fermi surface.}
\label{Fig1}
\end{figure}
%%%%%%%%%%%%%

In Fig.\ref{Fig1}(a) we illustrate the arrangement of spin-$1/2$ fermions in the limiting case
of zero temperature. In this regime and in accordance with the Pauli exclusion principle, 
particles occupy the lowest available single-particle states sequentially---one with spin up 
and the other one with spin down---until all particles have been placed. The highest occupied 
state defines the Fermi level, with the corresponding momentum and energy referred to as the 
Fermi momentum and Fermi energy, respectively. The $T\!=\!0$ curve in Fig.\ref{Fig1}(b) 
encapsulates this situation: at $T\!=\!0$ ($\beta\rightarrow\infty$), all states with a given spin 
and with energies below the chemical potential $\mu$ are completely filled while all the states 
above $\mu$ remain empty, resulting in a sharp Fermi surface.  However, at finite temperature 
($T\!\neq\!0$) thermal effects smear the Fermi surface by promoting some fermions below the
Fermi surface to states above it.

\subsection{A one-component Fermi gas}
\label{sec:OneComponent}

At zero temperature, the Fermi wavenumber $k_{\rm F}$ serves as a convenient proxy for the 
number density $n$ of the system. To determine the total number of particles $N$, one must sum 
over all occupied single-particle states, namely,
%%%%%%%%%%%%%
\begin{equation}
 N = 2\sum_{\bf k}\Theta(k_{\rm F}-|{\bf k}|) \to 
 2\int V\frac{d^{3}k}{(2\pi)^{3}}\Theta(k_{\rm F}-|{\bf k}|) = 
 \frac{V}{\pi^{2}}\int_{0}^{\,k_{\rm F}}\!\!k^{2}dk=V\frac{k_{F}^{3}}{3\pi^{2}}
 \implies n \equiv \frac{N}{V} = \frac{k_{F}^{3}}{3\pi^{2}} = \frac{p_{F}^{3}}{3\pi^{2}\hbar^{3}}.
 \label{Density} 
\end{equation}
%%%%%%%%%%%%%
Here the factor of $2$ reflects the spin degeneracy, $V$ denotes the volume of the system, 
$\Theta(x)$ is the Heaviside step function, and the first arrow indicates that in the thermodynamic 
limit---where both N and V approach infinity while their ratio remains finite---the discrete sum over 
states can be replaced by a continuous integral. The last expression determines the relation 
between the number density and the Fermi momentum $p_{\rm F}\!\equiv\!\hbar k_{F}$.

But why is the zero-temperature limit relevant, given that the core temperature of a young neutron 
star may reach a value as high as $T\!=\!10^{9}$K? At finite temperature, states just above the 
Fermi level begin to be populated at the expense of a corresponding depletion in the occupancy of 
states just below it; this situation is illustrated by the $T\!\ne\!0$ curve in Fig.\ref{Fig1}(b).  In this 
regime, only states within an energy interval of the order of  $k_{\rm B}T\!\approx\!0.1\,{\rm MeV}$  
around the Fermi level can be thermally excited to higher unoccupied states---those farther below 
remain inaccessible because they are Pauli blocked. The number of fermions excited above the 
Fermi level, that is, the approximate width of the $T\!\ne\!0$ curve, scales with the ratio of 
the temperature $T$ to the Fermi temperature $T_{\rm F}\!\equiv\!\varepsilon_{\rm F}/k_{\rm B}$.
Unlike the physical temperature, the Fermi temperature is entirely determined by the number 
density of the system. Under the extreme densities typical of a neutron-star core, this ratio is 
exceedingly small, yielding a fraction of thermally excited neutrons of about
%%%%%%%%%%%%%
\begin{equation}
 \frac{N_{\rm ex}}{N} \sim \frac{T}{T_{\rm F}} \approx 10^{-3},
 \label{NxoverN}
\end{equation}
%%%%%%%%%%%%%
indicating that only about one in a thousand neutrons participates in thermal processes 
within the neutron star, such as heat capacity and thermal conductivity. Even more 
suppressed by the small ratio $T/T_{\rm F}$, are the thermal contributions to the energy 
density and pressure of the star, both of which scale as $(T/T_{\rm F})^{2}\!\sim\!10^{-6}$. 
Accordingly, in modeling the structure of a neutron star, it is an excellent approximation 
to set $T\!\equiv\!0$. Hence, we now proceed to compute the equation of state (EOS) of 
a zero-temperature, relativistic free Fermi gas. 

At zero temperature, the EOS of a free Fermi gas of neutrons (or any fermion) is obtained 
by filling the lowest available single-particle states consistent with the Pauli exclusion principle, 
as depicted in Fig.\ref{Fig1}. Unlike the Fermi momentum---which is fixed solely by the number 
density and is therefore independent of the dispersion relation, the energy density and pressure 
of the system depend sensitively on the relation between the energy and the momentum. In 
particular, non-interacting fermions of rest mass $m$ obey the relativistic dispersion relation
%%%%%%%%%%%%%
\begin{equation}
 \varepsilon(p)\!=\!\sqrt{(pc)^{2}+(mc^{2})^{2}},
 \label{EvsP}
\end{equation}
%%%%%%%%%%%%%
where $c$ is the speed of light in vacuum and the momentum $p\!=\!\hbar k$ is simply related 
to the wave number $k$. Following similar steps as in Eq.(\ref{Density}), the energy density of 
a zero-temperature Fermi gas is given by
%%%%%%%%%%%%%
\begin{align}
   {\mathcal E}(n) & \equiv \frac{E}{V} = 2\int\frac{d^{3}k}{(2\pi)^{3}}\varepsilon(k)\Theta(k_{\rm F}-|{\bf k}|) = 
   \frac{1}{\pi^{2}}\int_{0}^{\,k_{\rm F}}\!\!\sqrt{(\hbar c k)^{2}+(mc^{2})^{2}}k^{2}dk \\
   & = \frac{(mc^{2})^{4}}{(\hbar c)^{3}}  \frac{1}{\pi^{2}}\int_{0}^{\mathlarger{x}_{\rm F}}x^{2}\sqrt{1+x^{2}}\,dx= 
   {\mathcal E}_{\mspace{2mu}0} \Big[\xFermi{}\yFermi{}(\xFermi{2}+\yFermi{2}) -\ln(\xFermi{}+\yFermi{})\Big],
 \label{EDensity} 
\end{align}
%%%%%%%%%%%%%
where ${\mathcal E}_{\mspace{2mu}0}$ sets the energy scale, and $\xFermi{}$ and $\yFermi{}$ are
the dimensionless Fermi momentum and Fermi energy:
%%%%%%%%
\begin{equation}
 {\mathcal E}_{\mspace{2mu}0} \equiv \frac{1}{8\pi^{2}}\frac{(mc^{2})^{4}}{(\hbar c)^{3}}, \hspace{8pt}
 \xFermi{} \equiv \frac{\pFermi{}\,c}{mc^{2}}, \hspace{8pt}
  \yFermi{} \equiv \frac{\eFermi{}}{mc^{2}}=\sqrt{1+\xFermi{2}}\,,
 \label{E0}
\end{equation}
%%%%%%%%%

Note that while the above expression for the energy density is exact, it is instructive to examine its behavior in 
both the nonrelativistic ($\xFermi\!\ll\!1$) and ultra-relativistic ($\xFermi\!\gg\!1$) limits. That is,
%%%%%%%%%%%%
\begin{equation}
{\mathcal E}{\mspace{2mu}(n)} \to
  n mc^{2}\begin{cases} 
   1 + \displaystyle{\frac{3}{10}}\xFermi{2} & \mbox{if\, } \xFermi{}\!\ll\!1,
   \vspace{3pt} \\ 
   \displaystyle{\frac{3}{4}\xFermi{}} & \mbox{if\, } \xFermi{}\!\gg\!1.
  \end{cases}    
\label{EFermiGas}
\end{equation}
%%%%%%%%%%%%

The $T\!=\!0$ equation of state relevant to the structure of compact stars, either white dwarfs or neutron stars, 
involves a relation between the pressure support against gravitational collapse and the energy density that is 
the source of gravity. In general, this relation is provided in parametric form; that is, both the pressure and the 
energy density are given in terms of the number density. In the context of a Fermi gas at zero temperature---even 
in the presence of interactions---the Hugenholtz--Van-Hove theorem\,\citep{Hugenholtz1958}---plays a fundamental 
role, as it connects three key physical quantities: the energy density, the pressure, and the Fermi energy. The 
theorem may be expressed as follows:
%%%%%%%%%%%%
\begin{equation}
 \frac{E}{N} + \frac{P}{n} = \eFermi{}  \iff {\mathcal E} +P = n\eFermi{}. 
 \label{HVH1}
\end{equation}
%%%%%%%%%%%%
Invoking the Hugenholtz--Van Hove theorem and following the same definitions as in Eq.\,(\ref{E0}), one obtains 
the pressure of a free Fermi gas:
%%%%%%%%%%%%
\begin{equation}
 P(n) = {\mathcal E}_{\mspace{2mu}0}
 \left[\displaystyle{\frac{2}{3}}{\xFermi{3}}\hspace{0.5pt}\yFermi{} 
     - \xFermi{}\hspace{0.5pt}\yFermi{}+\ln\left(\xFermi{}+\yFermi{}\right)\right]
     \rightarrow nmc^{2}
    \begin{cases} 
     \displaystyle{\frac{\xFermi{2}}{5}} & \mbox{if\, } \xFermi{}\!\ll\!1\,, 
     \vspace{5pt} \\ 
     \displaystyle{\frac{\xFermi{}}{4}} & \mbox{if\, } \xFermi{}\!\gg\!1\,,    
  \end{cases}     
\label{HVH2}
\end{equation}
%%%%%%%%%%%%
where the arrow in the last expression denotes the appropriate nonrelativistic and relativistic limits. 

In the particular case of white-dwarf stars---the final evolutionary stage of low-mass stars like our Sun---their stability 
and eventual gravitational collapse are both manifestations of physics beyond classical Newtonian theory. For 
low-mass white dwarfs, gravitational collapse is prevented by electron degeneracy pressure, which arises from the 
Pauli exclusion principle. However, as the mass of the star increases—and with it the electronic density—the electrons 
become relativistic, causing the pressure to soften from scaling as $P\sim n^{5/3}$ to $P\sim n^{4/3}$; see Eq.(\ref{HVH2}). 
This reduction in pressure is no longer sufficient to counteract gravity, and the star collapses once it reaches the 
Chandrasekhar mass, $M_{\rm ch}\!\approx\!1.4\,M_{\odot}$\,\citep{Chandrasekhar:1931}. Although unknown at the 
time, we now understand that collapsing white dwarfs form either neutron stars or black holes. Remarkably, when 
Chandrasekhar made this groundbreaking prediction as a newly minted graduate student, the neutron had not yet 
been discovered---Chadwick’s seminal paper on the “Possible Existence of a Neutron” appeared a year 
later\,\citep{Chadwick:1932}. Given that the pressure support in white-dwarf stars arises primarily from electron 
degeneracy pressure, it is instructive to solve the equations of hydrostatic equilibrium in this context before 
delving into the study of neutron stars\,\citep{Jackson:2004vt}.

\subsection{A two-component Fermi gas}
\label{sec:TwoComponent}

Although a neutron star is most commonly perceived as a conglomerate of neutrons, it is energetically advantageous 
for some of the neutrons near the Fermi surface to undergo beta decay into protons, electrons, and antineutrinos. In 
anticipation of this more general situation, we aim to compute the energy and pressure of a system composed of both 
neutrons and protons, under the assumption that the neutron and proton densities are each conserved individually; that is, 
we neglect electromagnetic as well as weak interactions. In this simplified scenario, the individual neutron $(n_{n})$ and 
proton $(n_{p})$ densities are treated as fixed, predetermined quantities. To quantify a possible neutron-proton asymmetry, 
we introduce a dimensionless asymmetry parameter $\alpha$ given by
%%%%%%%%%%%%
\begin{equation}
  \alpha \equiv \frac{n_{n} - n_{p}}{n_{n} + n_{p}} = \frac{n_{n} - n_{p}}{n}.
\label{alpha}
\end{equation}
%%%%%%%%%%%%
Following this convention, $\alpha\!=\!0$ corresponds to symmetric nuclear matter with equal numbers of neutrons 
and protons, while $\alpha\!=\!1$ represents pure neutron matter---a system made entirely of neutrons. As in the
previous section, the Fermi momentum plays an important role as a proxy for the density. For a two-component
system we must introduce individual neutron $\kFermi{n}$ and proton $\kFermi{p}$ Fermi momenta through the 
following definitions:
%%%%%%%%%%%%
\begin{subequations}
\begin{align}
  n_{n} & = \frac{\left(\kFermi{n}\right)^{3}}{3\pi^{2}} =
  \left(\frac{1+\alpha}{2}\right)n = (1+\alpha)\frac{\kFermi{3}}{3\pi^{2}}\,,\\
  n_{p} & = \frac{\left(\kFermi{p}\right)^{3}}{3\pi^{2}} =
  \left(\frac{1-\alpha}{2}\right)n = (1-\alpha)\frac{\kFermi{3}}{3\pi^{2}}.
\end{align}
\label{nnnp}
\end{subequations}
%%%%%%%%%%%%
Note that in this case the Fermi momenta $\kFermi{}$ has been defined as a proxy for the \emph{total} density 
of the system. This yields the following expressions for the individual Fermi momenta:
%%%%%%%%%%%%
\begin{equation}
n_{n} + n_{p} = n  = \frac{2k_{\text{F}}^3}{3\pi^{2}} \to
\begin{aligned}
  \;\; k_{\text{F}}^{n} &= (1+\alpha)^{1/3} k_{\text{F}},  \\
  \;\; k_{\text{F}}^{p} &= (1-\alpha)^{1/3} k_{\text{F}}.
\end{aligned}
\label{kFermi}
\end{equation}
%%%%%%%%%%%%

We are now in a position to compute the energy per particle of a two-component system. We anticipate 
that as a consequence of the Pauli exclusion principle, the energy will increase as the system develops 
a neutron-proton asymmetry. The Pauli exclusion principle plays a dominant role because, once the system 
acquires a neutron excess, protons occupying states near the Fermi surface must be promoted to unoccupied 
states above the neutron Fermi surface and this costs energy. This process is the isospin analogue of flipping 
a spin in Fig.\ref{Fig1}.

To provide a more intuitive picture, it is customary to address the role of a neutron-proton asymmetry on the 
energy of the system in terms of two main contributions: (a) the energy of symmetric ($n_{n}\!=\!n_{p}$) nuclear 
matter and (b) the symmetry energy that provides a correction to the symmetric limit to account for the cost 
of turning protons into neutrons (or vice versa). To do so, we provide a Taylor expansion of the energy per 
nucleon in even powers of $\alpha$:
%%%
\begin{equation}
 \frac{E}{A}(n,\alpha) =  \frac{E}{A}(n,\alpha\!\equiv\!0) + \alpha^{2}S(n) +  \alpha^{4}S_{4} (n) + \ldots 
\label{EAsym0}
\end{equation}
%%%
The reason that only even powers of $\alpha$ appear in the expansion is that, in the absence Coulomb 
interactions, it is equally costly to turn protons into neutrons than neutrons into protons. Note that throughout 
this contribution we neglect the very small neutron-proton mass difference. The first term in the above expansion 
represents the energy per particle of symmetric nuclear matter and may be directly obtained from the energy 
density listed in Eq.\,(\ref{EDensity}). That is,
%%%
\begin{equation}
  \frac{E}{A}(n,\alpha\!\equiv\!0) =  \frac{{\mathcal E}{\mspace{2mu}(n)}}{n}.
\label{Esym}
\end{equation}
%%%
The leading-order correction to the energy of symmetric nuclear matter is the symmetry energy $S(n)$, which
plays a critical role in the structure and dynamics of both neutron-rich nuclei and neutron stars; see
Refs.\citep{Tsang:2012se,Horowitz:2014bja,Lattimer:2014,Thiel:2019tkm,Mammei:2023kdf,BurgioFantina:2024},
and references contained therein.

Assuming, as we have done so far, that the only important correlations are those induced by the Pauli exclusion 
principle, one obtains a remarkably simple expression for the symmetry energy: 
%%%
\begin{equation}
 \frac{S(n)}{mc^{2}} = \frac{\xFermi{2}}{6\,\yFermi{}}  \rightarrow \frac{1}{6}
  \begin{cases} 
    \xFermi{2} & \mbox{if\, } \xFermi{}\!\ll\!1\,, 
   \vspace{3pt} \\ 
    \xFermi{} & \mbox{if\, } \xFermi{}\!\gg\!1\,.   
  \end{cases}  
 \label{STwo}
\end{equation}
%%%
That is, the energy cost of converting protons into neutrons (or vice versa) increases quadratically with the 
Fermi momentum at low densities and becomes linear at very high densities. However, given that the neutron 
excess in neutron-star matter is large under conditions of chemical equilibrium, it is important to assess whether 
neglecting higher-order terms in the Taylor expansion is justified. It has been shown in Ref.\,\citep{Piekarewicz2016}
that successive terms in the expansion are strongly suppressed relative to the preceding ones. Consequently, 
we conclude---at least for the case of a relativistic free Fermi gas---that the so-called parabolic approximation, 
namely,
%%%
\begin{equation}
 \frac{E}{A}(n,\alpha) \approx 
 \frac{E}{A}(n,\alpha\!\equiv\!0) + \alpha^{2}S(n) \,,  
 \label{EAsym1}
\end{equation} 
%%%
provides an excellent approximation to the energy of asymmetric matter. In particular, one can establish the 
following important connection between the energy of pure neutron matter ($\alpha\!=\!1$), the energy of 
symmetric nuclear matter ($\alpha\!=\!0$), and the symmetry energy:
%%%
\begin{equation}
   \frac{E}{A}(n,\alpha\!\equiv\!1) \approx
    \frac{E}{A}(n,\alpha\!\equiv\!0) + S(n) \,.
 \label{EPNM}
\end{equation} 
%%%
In this context, the symmetry energy may be interpreted as the energy cost required to convert symmetric 
nuclear matter into pure neutron matter. This intuitive picture remains valid in many---although not all---more 
realistic models.

\subsection{Chemical Equilibrium}
\label{sec:ChemEq}

We have now assembled all the ingredients required to construct the equation of state of neutron-star matter---at least 
within the Fermi-gas limit. As a macroscopic system containing a baryon number of the order of $10^{57}$, neutron-star 
matter must include leptons in order to ensure charge neutrality. Moreover, because neutron stars cool predominantly 
through neutrino emission via Urca processes, which involve both neutron beta decay and electron capture,
%%%%%%%%%%%
\begin{equation}
 n \rightarrow p + e^{-} + \bar{\nu}_{e}\; \leftrightarrow\; e^{-} + p \rightarrow n + \nu_{e},
\label{Urca}
\end{equation}
%%%%%%%%%%%
chemical equilibrium must also be enforced. Hence, assuming massless neutrinos with vanishing chemical potential---an 
excellent approximation for cold, fully catalyzed neutron stars---the energy density may be written as
%%%
\begin{equation}
 \frac{E}{V}(\rho,\alpha) =  \frac{{E}_{n}}{V}(\rho_{n}) + \frac{{E}_{p}}{V}(\rho_{p}) + \frac{{E}_{e}}{V}(\rho_{p}),
\label{EANstar}
\end{equation}
%%%
where, by charge neutrality, the electron and proton densities are equal. Moreover, to avoid confusion, we denote the baryon 
density by $\rho$ rather than by $n$. That is,
%%%%%%%%%%%%
\begin{equation}
  \rho_{n} = \frac{(1+\alpha)}{2}\rho \quad{\rm and}\quad
  \rho_{p} =\rho_{e} = \frac{(1-\alpha)}{2}\rho. 
 \label{Densities}
\end{equation}
%%%%%%%%%%%%

Unlike asymmetric nuclear matter, where both the baryon density $\rho$ and the neutron–proton asymmetry $\alpha$ may be 
treated as independent input parameters, in fully catalyzed neutron-star matter the asymmetry must be determined dynamically 
by minimizing the total energy density at each density. Specifically, the equilibrium value of $\alpha$ is obtained by solving the 
following transcendental equation:
%%%%%%%%%%%%
\begin{equation}
 \left(\frac{\partial E/V}{\partial\alpha}\right)_{\rho}=0 \rightarrow
  \left(\frac{\partial E_{n}}{\partial N}\right)_{V,T=0}\left(\frac{\partial\rho_{n}}{\partial\alpha}\right)_{\rho} +
  \left[\left(\frac{\partial E_{p}}{\partial Z}\right)+\left(\frac{\partial E_{e}}{\partial Z}\right)\right]_{V,T=0}
  \left(\frac{\partial\rho_{p}}{\partial\alpha}\right)_{\rho} = \frac{\rho}{2}\left[\mu_{n}-\mu_{p}-\mu_{e}\right]=0.
 \label{ChemEq0}
\end{equation}
%%%%%%%%%%%%
Thus, in chemical equilibrium, the neutron-proton asymmetry as a function of the baryon density $\rho$ is fixed by enforcing the 
condition
%%%%%%%%%%%%
\begin{equation}
  \mu_{n} = \mu_{p} + \mu_{e}.
 \label{ChemEq1}
\end{equation}
%%%%%%%%%%%%
This result is already implicit in Eq.(\ref{Urca}) upon assuming a vanishing neutrino chemical potential. Although the condition is 
entirely general for neutron-star matter, in the particular case in which neutrons, protons, and electrons may be treated as relativistic 
free Fermi gases one finds
%%%%%%%%%%%%
\begin{equation}
  \yFermi{n} = \yFermi{p} + \xFermi{p} \iff \frac{\rho_{p}}{\rho_{n}} 
                   = \frac{1}{8}\left(\frac{\xFermi{n}}{\yFermi{n}}\right)^{3} \rightarrow \frac{1}{8},
 \label{ChemEq2}
\end{equation}
%%%%%%%%%%%%
where the electron mass has been neglected and the arrow denotes the high-density limit, $m/\kFermi{n}\!\to\!0$. Thus, at 
sufficiently high densities, the ratio of protons to neutrons approaches $\rho_{p}/\rho_{n}\simeq 1/8$. Equivalently, the proton 
fraction tends to $Y_{p}=1/9$ and the neutron–proton asymmetry approaches $\alpha\!=\!7/9$. It is the requirement of charge 
neutrality that plays a dominant role in producing the large neutron excess characteristic of neutron stars.

\section{The Walecka Model}
\label{sec:Walecka}

The Walecka model\,\citep{Walecka:1974qa}, which will be the focus of this section, provides a relativistic 
framework for understanding the structure of atomic nuclei. Among the earliest attempts at a relativistic 
description of the nuclear dynamics are the works of Johnson and Teller\,\citep{Johnson:1955zz}, 
Duerr\,\citep{Duerr:1956zz}, and Miller and Green\,\citep{Miller:1972zza}. For a complete historical account---as 
well as a discussion of some of the early refinements to the Walecka model---see Ref.\citep{Serot:1984ey}. 
Beyond the desire to describe the properties of nuclear matter and their impact on ground-state energies 
and densities of atomic nuclei, a key motivation for a relativistic approach---and one that remains true to this 
day---was the development of a theory of highly condensed matter that remains causal at all densities and 
that can be applied to neutron stars\,\citep{Walecka:1974qa}.

%%%%%%%%%%%%%
\begin{figure}[h]
\centering
\includegraphics[width=0.35\textwidth]{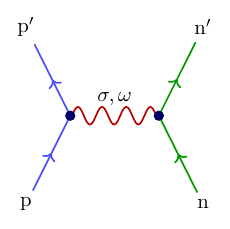}
\caption{Feynman diagram illustrating the nucleon-nucleon interaction in the Walecka ($\sigma$-$\omega$) 
              model.}
\label{Fig2}
\end{figure}
%%%%%%%%%%%%%

The need for a relativistic theory of atomic nuclei is not rooted in the idea that nucleons move at relativistic 
speeds; indeed, the dimensionless Fermi momentum defined in Eq.(\ref{E0}) is significantly less than one. 
Rather, the justification lies in the modest binding energy of nuclei that emerges from a strong cancellation 
between an intermediate-range attraction and a strong short-range repulsion. In this context, nucleons satisfy 
a Dirac equation in the presence of large scalar and vector potentials---a hallmark of the relativistic approach. 
This fact alone reproduces the highly successful phenomenology of the nuclear shell model: a relatively small 
binding energy accompanied by a strong spin-orbit potential\,\citep{Serot:1984ey}. Moreover, as we show below, 
the Walecka model provides a natural explanation for the saturation of symmetric nuclear matter.

In the original Walecka model---also known as the $\sigma$–$\omega$ model---nucleons interact through the 
exchange of a scalar $\sigma$ meson, which provides the intermediate-range attraction, and a vector $\omega$ 
meson, which generates the short-range repulsion. A representative Feynman diagram illustrating such a two-body 
interaction between a proton and a neutron is shown in Fig.~\ref{Fig2}.

Because our primary goal is to determine the equation of state of uniform nuclear matter, we assume the system 
to be both translationally and rotationally invariant. Under these conditions, the Lagrangian density in the mean-field 
approximation is given by
%%%%%%%%
\begin{eqnarray}
\mathscr{L}_{\rm MFT}=
\bar\psi \left[i\gamma^{\mu}\partial_{\mu}-g_{\rm v}\gamma^{\;0} V_{0}
                   -(M-g_{\rm s}\phi_{0})\right]\psi - \frac{1}{2}m_{\rm s}^{2}\phi_{0}^{2}
                   + \frac{1}{2}m_{\rm v}^{2}V_{0}^{2},
 \label{L1}
\end{eqnarray}
%%%%%%%%
where $\psi$ denotes the nucleon field and we employ the standard representation of the Dirac $\gamma$ 
matrices\,\citep{Peskin1995}. The intermediate-range attraction is mediated by a scalar meson of mass $m_{\rm s}$ 
and coupling strength $g_{\rm s}$, represented by the mean field $\phi_{0}$. Conversely, the short-range repulsion 
is generated by a vector meson of mass $m_{\rm v}$ and coupling strength $g_{\rm v}$, described by the mean field 
$V_{0}$. The energy--momentum tensor associated with this mean-field Lagrangian is given by
%%%
\begin{equation} 
  T_{\mu\nu} = i\bar{\psi}\gamma_{\mu}\partial_{\nu}\psi + 
  \left(\frac{1}{2}m_{\rm s}^{2}\phi_{0}^{2} - \frac{1}{2}m_{\rm v}^{2}V_{0}^{2}\right)g_{\mu\nu}.
  \label{Tmunu}
\end{equation} 
%%%

In the mean-field approximation adopted here, the meson fields satisfy a set of Yukawa equations 
of the following form:
%%%%%%%%%%%%
\begin{subequations}
\begin{align}
   & (\nabla^{2} - m_{\rm s}^{2})\,\phi_{0} = -g_{\rm s}\,\rho_{\rm s} 
   \rightarrow g_{\rm s}\phi_{0} \equiv \Phi_{0} = \left(\frac{g_{\rm s}^{2}}{m_{\rm s}^{2}}\right)\rho_{\rm s},
   \label{Sigma}\\
   & (\nabla^{2} - m_{\rm v}^{2})V_{0} = -g_{\rm v}\rho_{\rm v} 
   \rightarrow g_{\rm v}V_{0} \equiv W_{0} = \left(\frac{g_{\rm v}^{2}}{m_{\rm v}^{2}}\right)\rho_{\rm v}. 
\end{align}
\label{MFields}
\end{subequations}
%%%%%%%%%%%%
where $\rho_{\rm s}$ is the scalar density, $\rho_{\rm v}$ the conserved vector (or baryon) density, and the 
arrow denotes the appropriate limit for a translational/rotational invariant system. Note that we adopted the
standard ``nuclear-physics" convention in which the baryon density is denoted by $\rho$ rather than by $n$,
as in the previous sections. In turn, the nucleon field satisfies a Dirac equation with attractive scalar and 
repulsive vector potentials. That is,
%%%
\begin{equation} 
  \Big({\bm\alpha} \cdot {\bf p} + \beta(M-\Phi_{0}) + W_{0}\Big)\,{\cal U}({\bf p},\lambda)=
   \varepsilon(p){\cal U}({\bf p},\lambda)  \iff
    \Big({\bm\alpha} \cdot {\bf p} + \beta M^{\star}\Big)\,{\cal U}({\bf p},\lambda)=
   \varepsilon^{\star}(p){\cal U}({\bf p},\lambda),  
 \label{DiracEq}
\end{equation} 
%%%
where $\bm{\alpha}\!=\!\gamma^{0}\bm{\gamma}$, $\beta\!=\!\gamma^{0}$, and ${\cal U}({\bf p},\lambda)$ denotes 
a four-component Dirac spinor with momentum ${\bf p}$ and spin projection $\lambda$. The arrow in the above equation 
indicates that, for all practical purposes, the resulting expression is formally identical to the free Dirac equation for a nucleon 
with an effective mass $M^{\star}\!\equiv\!M\!-\!\Phi_{0}$ and an effective energy 
$\varepsilon^{\star}(p)\!\equiv\!\varepsilon(p)\!-\!W_{0}$. This result implies that, within the mean-field approximation, the 
dispersion relation for a nucleon propagating in uniform nuclear matter takes the familiar relativistic form
\begin{equation}
 \varepsilon(p)\!=\!\sqrt{p^{2}+M^{\star 2}}+W_{0},
\label{EstarvsP}
\end{equation}
which is formally identical to the free-particle expression, aside from the appearance of the effective mass $M^{\star}$ and 
the constant vector mean field $W_{0}$. Throughout the remainder of this work, we adopt natural units with $\hbar\!=\!c\!=\!1$. 

Following the formalism developed in Sec.\ref{sec:OneComponent}, the energy density of the system may be written as
%%%%%%%%%%%%
\begin{equation}
 {\mathcal E} \equiv T_{00} = \frac{1}{2}m_{\rm s}^{2}\phi_{0}^{2} - \frac{1}{2}m_{\rm v}^{2}V_{0}^{2} +
                                               \frac{\gamma}{2\pi^{2}}\int_{0}^{k_{\rm F}}k^{2}\varepsilon(k)dk =
                                               \frac{g_{\rm s}^{2}}{2m_{\rm s}^{2}}\rho_{\rm s}^{2}+
                                               \frac{g_{\rm v}^{2}}{2m_{\rm v}^{2}}\rho_{\rm v}^{2}+ 
                                               \frac{\gamma}{2\pi^{2}}\int_{0}^{k_{\rm F}}k^{2}\varepsilon^{\star}(k)dk,
 \label{EstarV}
\end{equation}
%%%%%%%%%%%%
where $\gamma$ is the spin-isospin degeneracy: $\gamma\!=\!4$ for symmetric nuclear matter and $\gamma\!=\!2$ 
for pure neutron matter. As in Sec.\ref{sec:OneComponent}, the Fermi momentum can also be used as a proxy for 
the conserved vector density, namely, 
%%%%%%%%%%%%%
\begin{equation}
 \rho\!\equiv\! \rho_{\rm v} \equiv \frac{A}{V} = \gamma\frac{\kFermi{3}}{6\pi^{2}}.
 \label{VectorDensity} 
\end{equation}
%%%%%%%%%%%%%
which can then be used to obtain the energy density of an interacting Fermi gas of mass $M^{\star}$ as a function
of density. That is,
\begin{equation}
  {\mathcal E}_{\rm F}^{\star}(\rho) = {\mathcal E}_{\mspace{2mu}0}^{\star} 
  \Big[\xFermi{$\star$}\yFermi{$\star$}
  \Big(\xFermi{$\star 2$}+\yFermi{$\star 2$}\Big) -
  \ln(\xFermi{$\star$}+\yFermi{$\star$})\Big],
 \label{EVstar}
\end{equation}
%%%%%%%%%%%%%
where ${\mathcal E}_{\mspace{2mu}0}^{\star}$ sets the energy scale for the problem, and $\xFermi{$\star$}$ 
and $\yFermi{$\star$}$ are the dimensionless Fermi momentum and Fermi energy defined as
%%%%%%%%
\begin{equation}
  {\mathcal E}_{\mspace{2mu}0}^{\star} \equiv \frac{\gamma\,M^{\star 4}}{16\,\pi^{2}}, \hspace{8pt}
  \xFermi{$\star$} \equiv \frac{\kFermi{}}{M^{\star}}, \hspace{8pt}
  \yFermi{$\star$} \equiv \sqrt{1+{\mathlarger x}_{\rm F}^{\star 2}}\,.
 \label{E0star}
\end{equation}
%%%%%%%%%
This yields a closed-form expression for the total energy density of the system as a function of the conserved 
baryon density $\rho$:
%%%%%%%%%%%%
\begin{equation}
 {\mathcal E}(\rho) = \frac{g_{\rm s}^{2}}{2m_{\rm s}^{2}}\rho_{\rm s}^{2}+
                                 \frac{g_{\rm v}^{2}}{2m_{\rm v}^{2}}\rho_{\rm v}^{2}+ 
                                 {\mathcal E}_{\mspace{2mu}0}^{\star} 
                                 \Big[\xFermi{$\star$}\yFermi{$\star$}
                                 \Big(\xFermi{$\star 2$}+\yFermi{$\star 2$}\Big) -
                                 \ln(\xFermi{$\star$}+\yFermi{$\star$})\Big].
 \label{EVTotal}
\end{equation}
%%%%%%%%%%%%
At first glance, it may appear that---once the conserved baryon density $\rho$ is specified---the energy density 
of the system could be obtained directly by evaluating the preceding expression. In practice, however, because
the scalar density is not conserved, one must  first determine the effective nucleon mass $M^{\star}$ as a function 
of the baryon density. Using Eq.(\ref{Sigma}), the effective mass satisfies the following transcendental equation:
%%%%%%%%%%%%
\begin{equation}
  M^{\star} = M - \left(\frac{g_{\rm s}^{2}}{m_{\rm s}^{2}}\right)\rho_{\rm s}(\kFermi{},M^{\star})=
                    M - \left(\frac{g_{\rm s}^{2}}{m_{\rm s}^{2}}\right)
                           \left(\frac{\gamma\,M^{\star 3}}{4\,\pi^{2}}\right)
                            \Big[\xFermi{$\star$}\yFermi{$\star$} -
                            \ln(\xFermi{$\star$}+\yFermi{$\star$})\Big].
 \label{Mstar}
\end{equation}
%%%%%%%%%%%%
Once the baryon density is fixed, this equation is solved self-consistently to obtain $M^{\star}$, which may then be 
substituted into Eq.(\ref{EVTotal}) to determine the energy density of the system at the given density.

In a similar fashion, one can determine the pressure of the system from the energy momentum tensor given
in Eq.(\ref{Tmunu}). That is,
%%%%%%%%%%%%
\begin{equation}
 P = \frac{1}{3}T_{ii} = - \frac{g_{\rm s}^{2}}{2m_{\rm s}^{2}}\rho_{\rm s}^{2}+
                                      \frac{g_{\rm v}^{2}}{2m_{\rm v}^{2}}\rho_{\rm v}^{2}+ 
   \frac{\gamma}{6\pi^{2}}\int_{0}^{k_{\rm F}}\frac{k^{4}}{\varepsilon^{\star}(k)}dk,                                 
 \label{Pstar0}
\end{equation}
%%%%%%%%%%%% 
or evaluating the integral in closed form, one obtains:
\begin{equation}
 P(\rho) = - \frac{g_{\rm s}^{2}}{2m_{\rm s}^{2}}\rho_{\rm s}^{2}+
                   \frac{g_{\rm v}^{2}}{2m_{\rm v}^{2}}\rho_{\rm v}^{2}+                                                
       {\mathcal E}_{\mspace{2mu}0}^{\star} \Big[\frac{2}{3}\xFermi{$\star$3}\yFermi{$\star$} - 
       \xFermi{$\star$}\yFermi{$\star$} + \ln(\xFermi{$\star$}+\yFermi{$\star$})\Big].                                      
 \label{Pstar1}
\end{equation}
%%%%%%%%%%%%
As in the case of the energy density, one is able to compute the pressure at a given density only after solving
the transcendental equation for the effective mass $M^{\star}$.

Finally, in neutron-star matter one must enforce chemical equilibrium as expressed in Eq.(\ref{ChemEq1}). In the 
Walecka model, where the vector interaction is purely isoscalar, the condition of chemical equilibrium leads to a 
relation that is nearly identical to that obtained for a free relativistic Fermi gas in Eq.(\ref{ChemEq2}), with the sole 
modification that the free nucleon mass is replaced by its effective mass, $M \to M^{\star}$. That is,
%%%%%%%%%%%%
\begin{equation}
  \yFermi{n$\star$} = \yFermi{p$\star$} + \xFermi{p$\star$} \iff \frac{\rho_{p}}{\rho_{n}} 
                   = \frac{1}{8}\left(\frac{\xFermi{n$\star$}}{\yFermi{n$\star$}}\right)^{3} 
                   \rightarrow \frac{1}{8}.
 \label{ChemEqStar}
\end{equation}
%%%%%%%%%%%%
Thus, as in the free Fermi–gas case, the proton fraction approaches $Y_{p}\!=\!1/9$ in the high-density limit, corresponding 
to a neutron-proton asymmetry of $\alpha\!=\!7/9$.

\section{Results}
\label{sec:Results}

As we have shown in Eq.(\ref{EAsym0}), the energy of a two-component system may be approximated as the 
sum of the energy of symmetric nuclear matter plus the symmetry energy---with the leading-order correction 
scaling as the fourth power of the neutron proton asymmetry $\alpha$. Using this expression, the rigorous
definition of the symmetry energy is given by
%%%%%%%%%%
\begin{equation}
 S(\rho) = \frac{1}{2}\!\left(\frac{\partial^{2}\mathlarger{\mathlarger{\varepsilon}}}
           {\partial\alpha^{2}}\right)_{\!\alpha=0}.
 \label{SymE0}
\end{equation}
%%%%%%%%%%

However, as argued earlier in the case of a free Fermi gas, the parabolic approximation is often used in the
literature to approximate the symmetry energy as the difference between the energy per nucleon of pure 
neutron matter and that of symmetric nuclear matter; that is,
%%%%%%%%%%
\begin{equation}
  S(\rho) \approx  \frac{E}{A}(\rho,\alpha\!\equiv\!1) -  \frac{E}{A}(\rho,\alpha\!\equiv\!0),
 \label{SymE1}
\end{equation} 
%%%%%%%%%%
Hence, within the scope of the parabolic approximation, the symmetry energy has a very simple interpretation:
it is the energy cost at fixed density of converting symmetric nuclear matter with equal numbers of protons and 
neutrons to pure neutron matter. In the context of a free Fermi gas, this cost is entirely dictated by the Pauli
exclusion principle. In the case of the Walecka model, the symmetry energy becomes a simple generalization 
of Eq.(\ref{STwo}), namely,  
%%%%%%%%%%
\begin{equation}
 S(\rho)=  M \frac{\xFermi{2}}{6\,\yFermi{}}  \rightarrow 
                M^{\star}\frac{\xFermi{$\star 2$}}{6\,\yFermi{$\star$}}.
  \label{SymE2}
\end{equation}
%%%%%%%%%%

%%%%%%%%%%%%%
\begin{figure}[h]
\centering
\includegraphics[width=0.7\textwidth]{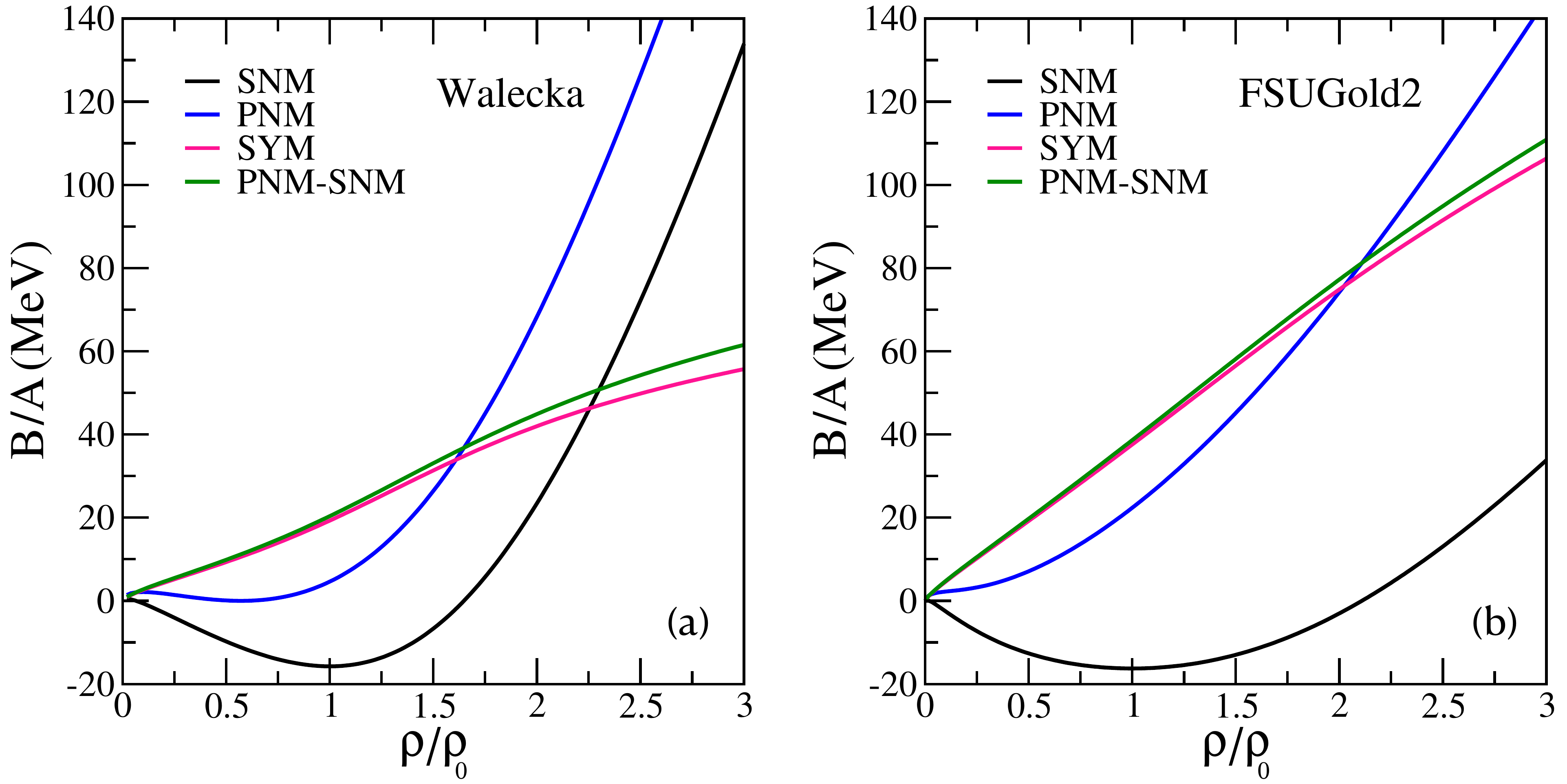}
\caption{Binding energy per nucleon for symmetric nuclear matter (SNM) and for pure neutron matter (PNM)
             as predicted by (a) the Walecka model\,\citep{Walecka:1974qa,Serot:1984ey} and by (b) the refined 
	     FSUGold2 model\,\citep{Chen:2014sca}. Also shown is the symmetry energy (SYM) and its 
	     parabolic approximation, defined as the energy per nucleon of pure neutron matter relative to 
	     corresponding quantity for symmetric nuclear matter (PNM-SNM).}
\label{Fig3}
\end{figure}
%%%%%%%%%%%%%

In Fig.\ref{Fig3}(a) we display the equation of state for both symmetric nuclear matter (SNM) and pure neutron 
matter (PNM) as predicted by the Walecka model\,\citep{Walecka:1974qa,Serot:1984ey}. The binding energy 
per nucleon is defined by subtracting the nucleon rest mass from the total energy per particle, namely,
$B/A\!=\!E/A\!-\!M$. We note that SNM saturates: the binding energy per nucleon reaches a minimum at a 
density $\rho_{0}$, which implies that the pressure vanishes at saturation. Saturation is a hallmark of the 
nuclear dynamics which manifests itself in the nearly constant interior density of heavy nuclei. In the particular 
case of the Walecka model, the saturation mechanism is especially instructive. Although the scalar attraction 
is stronger than the corresponding vector repulsion,
%%%%%%%%%%
\begin{equation}
 g_{\rm s}^{2}\frac{M^{2}}{m_{\rm s}^{2}} = 357.469 > 273.871 = g_{\rm v}^{2}\frac{M^{2}}{m_{\rm v}^{2}},
\label{Saturation}
\end{equation}
%%%%%%%%%%
the scalar attraction itself saturates. Specifically, the ratio of the scalar density to the conserved vector density 
vanishes at high densities. As a result, the scalar attraction becomes ineffective at sufficiently large densities, 
so the vector repulsion ultimately dominates. In contrast, pure neutron matter---with a spin-isospin degeneracy 
of $\gamma\!=\!2$ rather than $\gamma\!=\!4$ as in the case of symmetric nuclear matter---does not saturate. 

Also shown in Fig.\ref{Fig3}(a) is the symmetry energy computed both exactly and within the parabolic approximation, 
as given in Eqs.(\ref{SymE0}) and (\ref{SymE1}), respectively. While the parabolic approximation provides a reasonable 
description of the symmetry energy, it is not exact, particularly at high densities. Consequently, caution must be 
exercised when employing the parabolic approximation to quantify the role of the symmetry energy, especially in 
applications to neutron-star structure, where supranuclear densities are probed.

Since its inception more than five decades ago, the Walecka model\,\citep{Walecka:1974qa} has been systematically
 refined to achieve better agreement with both terrestrial experiments and astrophysical observations. One such 
 refinement led to the development of the FSUGold2 model, which was calibrated using the latest experimental data 
 and astrophysical constraints; see Ref.\,\citep{Chen:2014sca} and references contained therein. Although differences 
 in the predicted trends relative to the original Walecka model are already apparent in Fig.\ref{Fig3}(b), we conclude 
 this section by quantifying several of the most important distinctions between the two models.

To do so, we observe that in the vicinity of the saturation density $\rhozero$, the density dependence of both the 
energy per nucleon of symmetric nuclear matter as well as the symmetry energy are contained in a few bulk
parameters. That is,
%%%
\begin{subequations}
\begin{align}
 & \mathlarger{\varepsilon}_{{}_{\rm SNM}}(\rho) \equiv \mathlarger{\varepsilon}(\rho,\alpha\equiv0)
      = \epszero + \frac{1}{2}K_{0}\,x^{2}+\ldots \\
 & {S}(\rho) = J + L\,x + \ldots 
 %\frac{1}{2}K_{\rm sym}\,x^{2}\ldots   
\end{align} 
\label{EandS}
\end{subequations}
%%%
where $x\!=\!(\rho-\rhozero)\!/3\rhozero$ is a dimensionless parameter that quantifies the deviations 
of the density from its value at saturation. Because symmetric nuclear matter saturates, the expansion 
contains no term linear in $x$; that is, the pressure of symmetric nuclear matter vanishes at saturation. 
The leading-order correction is therefore governed by the incompressibility coefficient $K_{0}$, which 
has been constrained by measurements of giant monopole resonances\,\citep{Garg:2018uam} to lie in 
the interval $K_{0}\!\sim\!220$–$240\mathrm{\,MeV}$---in excellent agreement with the value predicted
by the FSUGold2 model, as shown in Table\,\ref{Table1}. Since the two free parameters of the Walecka 
model are calibrated to reproduce the saturation density and the binding energy per nucleon at saturation  
of symmetric nuclear matter, the resulting prediction for the incompressibility coefficient $K_{0}$ is a
genuine prediction, albeit one that lies well outside the experimentally determined range; see 
Table\,\ref{Table1}.

%%%%%%%%%%%%%%%%%%%%
\begin{table}[h]
\begin{center}
\begin{tabular}{|l||c|c|c|c|c|c|c|}
\hline\rule{0pt}{2.5ex}   
\!\!Model   &  $\rhozero$  &  $\epszero$  &  $M^{\star}/M$ & $K_{0}$  &  $J$  &  $L$ \\
\hline
\hline
Walecka\,\citep{Walecka:1974qa,Serot:1984ey}   & 0.149 & -15.75  & 0.541 & 547.16 & 19.29  & 68.46      \\
FSUGold2\,\citep{Chen:2014sca}                          & 0.150 &  -16.27 & 0.593 & 237.88 &  37.59 & 112.72     \\ 
\hline
\end{tabular}
\caption{Bulk parameters of infinite nuclear matter at saturation density $\rhozero$ as predicted by both the 
Walecka and FSUGold2 models.The quantities $\epszero$,  $M^{\star}/M$, and $K_{0}$ represent the binding 
energy per nucleon, the effective mass, and incompressibility coefficient of symmetric nuclear matter, whereas 
$J$ and $L$ denote the energy and slope of the symmetry energy---all evaluated at $\rhozero$. With the exception 
of the saturation density that is given in ${\rm fm}^{-3}$ and $M^{\star}/M$ which is dimensionless, all other quantities 
are expressed in MeV.}
\label{Table1}
\end{center}
\end{table}
%%%%%%%%%%%%%%%%%%%%

Also listed in Table\,\ref{Table1} are the symmetry energy $J$ and its slope $L$ evaluated at saturation density. In the 
Walecka model, the symmetry energy arises entirely from the Pauli exclusion principle, since converting protons into 
neutrons (or vice versa) is Pauli blocked. In contrast, the FSUGold2 model includes additional dynamical contributions 
through the explicit coupling to the $\rho$ meson, the isovector partner of the $\omega$ meson. As illustrated in 
Fig.\,\ref{Fig3}, the $\rho$-meson contribution enhances the symmetry energy, leading to the values reported in 
Table\,\ref{Table1}. The significantly larger values predicted by the FSUGold2 model are in good agreement with the 
neutron-skin thickness of ${}^{208}$Pb extracted by the PREX collaboration~\citep{Adhikari:2021phr,Reed:2021nqk}.

\section{Conclusions and Outlook}
\label{sec:Conclusions}

In closing, we emphasize that the equation of state of infinite nuclear matter provides a unique window into the complex 
dynamics of the nuclear many-body system. The bulk properties of symmetric nuclear matter encode saturation, namely 
the existence of an equilibrium density that is reflected in the interior density of heavy nuclei. Moreover, unlike electronic 
systems, nuclear matter---composed of two distinct constituents, neutrons and protons---exhibits particularly rich dynamics 
as one moves away from the symmetric limit. Indeed, the symmetry energy, which quantifies the energetic cost of converting 
symmetric nuclear matter into pure neutron matter, lies at the forefront of the scientific mission of modern radioactive-beam 
facilities. Perhaps most compelling, however, is the central role played by the nuclear matter equation of state in shaping the 
structure, dynamics, and composition of neutron stars in the emerging era of multi-messenger astronomy. In this context, 
the microphysics governing neutron stars is ultimately encoded in the equation of state of nuclear matter under extreme 
conditions of density and neutron-proton asymmetry.

\begin{figure}[h]
\centering
\includegraphics[width=0.5\textwidth]{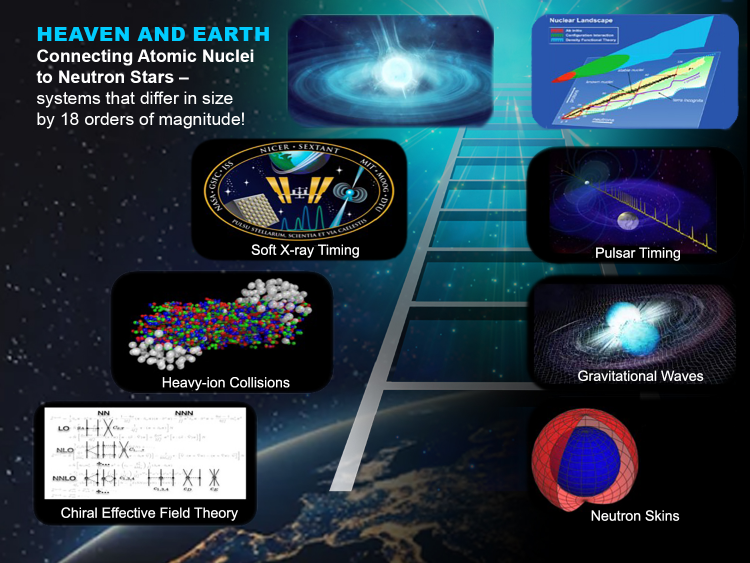}
\caption{Neutron-star inspired density ladder. Akin to the cosmological distance ladder, the confluence of so many advances 
in nuclear astrophysics motivates the creation of the equation of state density ladder---connecting theoretical models, terrestrial 
nuclear experiments, and astrophysical observations. As illustrated in the figure, no single method can determine the equation 
of state over its vast density domain, yet each rung on the ladder informs the equation of state in a suitable density domain
that overlaps with its neighboring rungs\,\citep{LRP2023}.}
\label{Fig4}
\end{figure}

The critical role of the nuclear matter equation of state can be made quantitative through a small set of bulk parameters 
defined at saturation density. Among these, the incompressibility of symmetric nuclear matter $K_{0}$ determines the 
stiffness of the equation of state near saturation and is constrained primarily by measurements of isoscalar giant monopole 
resonances in finite nuclei. In contrast, the density dependence of the symmetry energy, commonly characterized by its 
slope parameter $L$, controls the pressure of neutron-rich matter around saturation and plays a central role in 
determining the properties of both neutron-rich nuclei and neutron stars. At higher densities, relevant to the inner cores 
of neutron stars, the behavior of the equation of state is more naturally described in terms of the speed of sound
$dP/d{\mathcal E}$\,, whose density dependence encodes the microphysics that constrains the maximum mass, compactness, 
and possible phase structure of dense matter. Together, these parameters provide a coherent framework for connecting 
theoretical models, laboratory experiments, and astrophysical observations---an approach now commonly referred to as 
the density ladder, as illustrated in Fig.\,\ref{Fig4}\,\citep{LRP2023}.

%%%%%%%%%%%%%
%%%%%%%%%%%%%}

Within this broader landscape, the aim of this work was to present an introductory article accessible to a broad audience---including 
undergraduate students---that provides an overview of the wide-ranging applications of relativistic mean-field theory, with particular 
emphasis on the role of nuclear physics in constructing an equation of state capable of describing the remarkable properties of neutron 
stars. Rather than beginning with highly sophisticated models, we first established the essential concepts underlying the equation of 
state in the simple context of free Fermi gases. With these central ideas in place, we then moved to a more realistic 
description based on the relativistic Walecka model, in which the saturation properties of symmetric nuclear matter emerge naturally 
from the interplay of scalar and vector Lorentz fields. Finally, in light of the substantial progress achieved over the five decades since 
the introduction of the Walecka model, we introduced the FSUGold2 parametrization---a modern relativistic model accurately calibrated, 
with quantified theoretical uncertainties, to a broad set of laboratory and astrophysical observables---thereby mitigating many of the 
shortcomings of the original formulation.

%%%%%%%%%%%%%
\begin{figure}[h]
\centering
\includegraphics[width=0.45\textwidth]{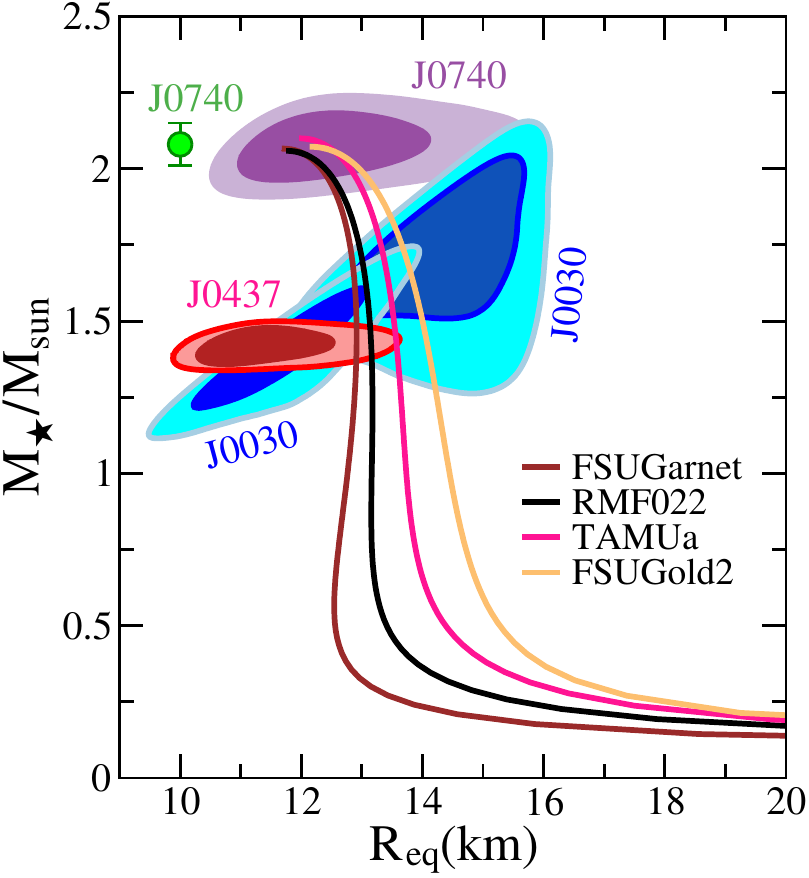}
\caption{Mass-radius relation for neutron stars as predicted by a set of four accurately calibrated relativistic 
models\,\citep{Fattoyev:2013yaa,Chen:2014sca,Chen:2014mza}. Also shown are NICER 68\% and 95\% 
credible intervals for the three millisecond pulsars: PSR J0030+0451\,\citep{Miller:2019cac,Riley:2019yda}, 
PSR J0740+6620\,\citep{Miller:2021qha,Riley:2021pdl}, and PSR J0437-4715\,\citep{Choudhury:2024xbk}.
The data point at $M_{\star}\!=\!(2.08\pm 0.07)\,M_{\odot}$ denotes the mass of PSR J0740+6620 previously
obtained via Shapiro delay\,\citep{Cromartie:2019kug,Fonseca:2021wxt}.}
\label{Fig5}
\end{figure}
%%%%%%%%%%%%%

Looking ahead, we conclude by illustrating in Fig.\,\ref{Fig5} recent progress in both astrophysical observations 
of neutron stars and theoretical developments within relativistic mean-field models. Details of the four theoretical 
models shown in the figure may be found in Refs.\,\citep{Fattoyev:2013yaa,Chen:2014sca,Chen:2014mza}. 
In turn, the three confidence contours represent simultaneous mass-radius measurements of the millisecond pulsars 
(MSPs) PSR~J0030+0451\,\citep{Miller:2019cac,Riley:2019yda}, PSR~J0740+6620\,\citep{Miller:2021qha,Riley:2021pdl}, 
and PSR~J0437-4715\,\citep{Choudhury:2024xbk}. This classification makes MSPs ideal targets for precision mass-radius 
measurements, one of the central goals of the NICER mission.

All three NICER targets are recycled neutron stars that have been spun up through accretion and possess relatively 
weak magnetic fields, which mitigate systematic uncertainties in modeling X-ray emission from localized hot spots on 
the stellar surface. Moreover, when part of binary systems, independent radio-timing mass measurements via Shapiro 
delay provide precise mass constraints that help break mass-radius degeneracies inherent to pulse-profile modeling. 
Such a mass measurement is also shown in Fig.\,\ref{Fig5} for PSR~J0740+6620, a neutron star that currently holds 
the record as the most massive---and best measured---neutron star to date\,\citep{Cromartie:2019kug,Fonseca:2021wxt}. 
As such, MSPs are uniquely suited for probing the equation of state of cold, dense matter. These developments---together 
with the historic detection of gravitational waves from neutron-star mergers\,\citep{Abbott:PRL2017}---have spearheaded 
the brand-new era of multi-messenger astronomy, placing nuclear science at the forefront of such an exciting and rapidly 
evolving field. We hope that this chapter will inspire a new generation of scientists and offer a clear entry point to the 
extensive and rapidly evolving literature on this fascinating topic.

%\bibliography{./ReferencesJP}
\bibliography{./Main.bbl}

\begin{ack}[Acknowledgments]
 \,This material is based upon work supported by the U.S. Department of Energy Office of Science, 
 Office of Nuclear Physics under Award Number DE-FG02-92ER40750.
\end{ack}

\end{document}